\documentclass{ws-procs10x7}

\begin{document}

\newcommand{\be}{\begin{equation}}
\newcommand{\ee}{\end{equation}}
\newcommand{\bea}{\begin{eqnarray}}
\newcommand{\eea}{\end{eqnarray}}

\title{QCD and String Theory 
}

\author{Igor R. Klebanov}

\address{Joseph Henry Laboratories, Princeton University, Princeton, NJ 08544,
USA\\E-mail: klebanov@princeton.edu}

\twocolumn[\maketitle\abstract{
This talk begins with some history and basic facts about string theory and
its connections with strong interactions. Comparisons
of stacks of Dirichlet branes
with curved backgrounds produced by them are used to motivate the AdS/CFT
correspondence between superconformal gauge theory and
string theory on a product of Anti-de Sitter space and a compact manifold.
The ensuing duality between semi-classical spinning strings
and long gauge theory operators is briefly reviewed.
Strongly coupled thermal SYM theory is explored via
a black hole in 5-dimensional AdS space, which 
leads to explicit results for its
entropy and shear viscosity. 
A conjectured universal lower bound on the viscosity to 
entropy density
ratio, and its possible relation to recent results from RHIC, are
discussed. Finally, some available results on string duals of confining
gauge theories are briefly reviewed. 
}]

\section{Introduction}

String theory 
\footnote{Due to a strict length limit,
I did not include figures in this manuscript. The figures are included in
the PowerPoint version of this talk, available at the 2005
Lepton-Photon Symposium web site.} 
is well known to be the leading prospect for 
quantizing gravity and
unifying it with other interactions\cite{Green,Polchinski}. 
One may also take a broader view of string
theory as a description of string-like excitations that 
arise in many different
physical systems, such as the superconducting 
flux tubes, cosmic strings, and
of course the chromo-electric flux tubes in non-Abelian gauge theories,
which are the subject of my talk.
You could object that these string-like excitations are ``emergent''
rather than fundamental phenomena. 
We will see, however, that there is no sharp
distinction between ``emergent'' and fundamental strings. We will
exhibit examples, stemming from the 
AdS/CFT correspondence\cite{jthroat,US,EW},
where the ``emergent'' and fundamental strings are
dual descriptions of the same theory. Besides being of great theoretical
interest, such gauge/string dualities are becoming a useful tool
for studying strongly coupled gauge theories. A developing connection
that is highlighted in this talk is with the new results at RHIC:\cite{Adcox} 
there are indications that a rather
strongly coupled Quark-Gluon Plasma (sQGP) 
has been observed.

\section{Some early history}\label{subsec:prod}

String Theory was born out of attempts to understand
	the Strong Interactions.
	Empirical evidence for a string-like structure of hadrons 
comes from arranging mesons and baryons into approximately
linear Regge trajectories. Studies of  
$\pi N$ scattering prompted
    Dolen, Horn and Schmid\cite{Dolen} to make a duality 
conjecture stating that 
the sum over s-channel exchanges 
equals the sum over t-channel ones. This posed the problem of finding
the analytic form of such dual amplitudes.
Veneziano\cite{Veneziano} found the first, and very simple, expression for a
 manifestly dual 4-point amplitude:
\be 
A(s,t) \sim {\Gamma (-\alpha(s))
\Gamma (-\alpha(t))\over
\Gamma (-\alpha(s) - \alpha(t)) }
\ee
with an exactly linear Regge trajectory
$ \alpha(s) = \alpha(0) + \alpha' s$.
Soon after, Nambu\cite{Nambu},
Nielsen\cite{Nielsen} and Susskind\cite{Susskind} 
independently proposed its open string 
interpretation. This led to an explosion of interest in the early 70's
in string
theory as a description of strongly interacting particles.
The basic idea is to think of a meson as an open string
with a quark at one end-point and an anti-quark at another. Then various
meson states arise as different excitations of such an open string. 
The splitting of a string describes the decay of a meson into two mesons,
for example.

The string world sheet dynamics is governed by the Nambu-Goto area action
\be S_{\rm NG}= -T \int d\sigma d\tau
\sqrt{-{\rm det}\ \partial_a X^\mu \partial_b X_\mu}
\ ,
\ee
where the indexes $a,b$ take two values ranging over the
$\sigma$ and $\tau$ directions on the world sheet.
The string tension is related to the Regge slope through
$ T={1\over 2\pi\alpha'} $.
The quantum consistency of the Veneziano model requires  that the Regge 
intercept is $\alpha(0)=1$,
so that the spin 1 state is massless but the spin 0 is a tachyon. 
But the $\rho$ meson is certainly not massless, and there are no
tachyons in the real world. This is how
the string theory of strong interactions started
to run into problems.

Calculation of the string zero-point energy gives
\be
\alpha(0) = {d-2\over 24}
\ .
\ee
Hence the model has to be defined in 26 space-time  dimensions.
Attempts to quantize such a string model directly in 3+1 dimensions led to 
tachyons and problems with unitarity.
Consistent supersymmetric string theories were discovered in 10 
dimensions, but their relation to the strong interactions was 
initially completely unclear.
Most importantly, the Asymptotic Freedom of strong interactions was 
discovered\cite{GWP}, 
singling out Quantum Chromodynamics (QCD) as the exact 
field theory of strong interactions.
At this point
most physicists gave up on strings as a description of strong interactions. 
Instead, since 
the graviton appears naturally in the closed string 
spectrum,       
string theory emerged as the leading hope for unifying 
quantum gravity with other forces\cite{Scherk}. 

\section{ QCD gives strings a chance }

Now that we know that a non-Abelian gauge theory is an exact
description of strong interactions, is there any room left for
string theory in this field? Luckily, the answer is positive.
At short distances, much smaller than 1 fermi, 
the quark anti-quark potential is Coulombic, due to 
Asymptotic Freedom.
At large distances the potential should be linear  
due to formation of a confining flux tube\cite{Wilson}.
When these tubes are much longer than their thickness, one can hope to
describe them, at least
approximately, by semi-classical Nambu strings\cite{Nambunew}.
This appears to explain the existence of approximately linear
Regge trajectories:
a linear relation between angular momentum and mass-squared 
\be J =\alpha' m^2 + \alpha (0)
\ ,
\ee
is provided 
by a semi-classical spinning relativistic string with massless quark 
and anti-quark at its endpoints.
A semi-classical string approach to the QCD flux tubes 
is widely used, for example, 
in jet hadronization algorithms based on the 
Lund String Model\cite{Andersson}. 

Semi-classical quantization around a long straight 
Nambu string predicts the quark anti-quark 
potential\cite{Luscher:1980fr}
\be
V(r) = Tr + \mu +{\gamma\over r} + O(1/r^2)
\ .\ee
The coefficient $\gamma$ 
of the universal L\" uscher term depends only on the 
space-time dimension $d$ and is proportional to the Regge intercept:
$\gamma=-\pi (d-2)/24$.
Recent lattice calculations of the force vs. distance for
probe quarks and anti-quarks\cite{Luscher} produce good 
agreement with this value in $d=3$ and $d=4$ 
for $r>0.7 fm$.
Thus, long QCD strings appear to be well described by the Nambu-Goto
area action. But quantization of short, highly quantum QCD strings,
that could lead to a calculation of light meson and glueball spectra,
is a much harder problem.

The connection of gauge theory with string theory is strengthened by
`t Hooft's generalization of QCD from 3 colors 
($SU(3)$ gauge group) to $N$ colors ($SU(N)$ gauge group)\cite{GT}.
The idea is to make $N$ 
large, while keeping the `t Hooft coupling 
$\lambda = g_{\rm YM}^2 N$ fixed.
In this limit each Feynman graph carries
a topological factor $N^\chi$, where $\chi$ is the Euler
characteristic of the graph.
Thus, the sum over graphs of a given topology can perhaps be thought of
as a sum over world sheets of a hypothetical ``QCD string.''
Since the spheres (string tree diagrams)
are weighted by $N^2$, the tori (string one-loop diagrams) --
by $N^0$, etc., we find that the closed string coupling constant is of order
$N^{-1}$. Thus, the advantage of taking $N$ to be large is that we find
a weakly coupled string theory. 
In the large $N$ limit the gauge theory 
simplifies in that 
only the planar diagrams contribute. But 
directly summing even this subclass of diagrams seems to be an
impossible task.
From the dual QCD string point of view, it
is not clear how to describe
this string theory in elementary terms.

Because of the difficulty of these problems,
between the late 70's and the
mid-90's many theorists gave up hope of finding an exact gauge/string duality.
One notable exception is Polyakov who already in 1981 proposed that 
the string theory dual to a 
4-d gauge theory should have a 5-th hidden dimension\cite{Polyakov}. 
In later work\cite{Polyakovnew} he refined this proposal, 
suggesting that the 5-d metric must
be ``warped.''

\section{The Geometry of Dirichlet Branes}

In the mid-nineties the Dirichlet branes,
or D-branes for short,
 brought string theory back to 
gauge theory.
The D-branes are soliton-like ``membranes'' of various internal dimensionalities
contained in theories of closed superstrings\cite{Polchinski}.
A Dirichlet $p$-brane (or D$p$-brane) is a $p+1$ dimensional hyperplane
in $9+1$ dimensional space-time where strings are allowed to end.
A D-brane is much like a topological defect: upon 
touching a D-brane, a closed string
can open up and turn into an open string whose
ends are free to move along the D-brane. For the end-points of such a string
the $p+1$ longitudinal coordinates satisfy the conventional free (Neumann)
boundary conditions, while the $9-p$ coordinates transverse to
the D$p$-brane have the fixed (Dirichlet) boundary conditions; hence
the origin of the term ``Dirichlet brane.'' 
In a seminal paper\cite{Polch} Polchinski
showed that a D$p$-brane 
preserves
$1/2$ of the bulk supersymmetries and carries an elementary unit
of charge with respect to the $p+1$ form gauge potential from the
Ramond-Ramond sector of type II superstring. 

For our purposes, the most important property of D-branes 
is that they realize
gauge theories on their world volume. The massless spectrum of open strings
living on a D$p$-brane is that of a maximally supersymmetric $U(1)$
gauge theory in $p+1$ dimensions. The $9-p$ massless
scalar fields present in this supermultiplet are the expected Goldstone 
modes associated with the transverse oscillations of the D$p$-brane,
while the photons and fermions
provide the unique supersymmetric
completion.
If we consider $N$ parallel D-branes,
then there are $N^2$ different species of open strings because they can
begin and end on any of the D-branes. 
$N^2$ is the dimension of the adjoint representation of $U(N)$,
and indeed we find the maximally supersymmetric $U(N)$ 
gauge theory in this setting.

The relative separations of the D$p$-branes in the $9-p$ transverse
dimensions are determined by the expectation values of the scalar fields.
We will be 
interested in the case where all scalar expectation
values vanish, so that the $N$ D$p$-branes are stacked on top of each other.
If $N$ is large, then this stack is a heavy object embedded into a theory
of closed strings which contains gravity. Naturally, this macroscopic
object will curve space: it may be described by some classical metric
and other background fields.
Thus, we have two very different descriptions of the stack of D$p$-branes:
one in terms of the $U(N)$ supersymmetric gauge theory on its world volume,
and the other in terms of the classical
charged $p$-brane background of the type II
closed superstring theory. The relation between these two descriptions
is at the heart of the 
connections between gauge fields and strings that are the subject of
this talk.

\subsection {Coincident D3-branes}

Parallel
D3-branes realize a $3+1$ dimensional $U(N)$ gauge theory,
which is a maximally supersymmetric ``cousin'' of QCD.
Let us compare a stack of D3-branes with the 
Ramond-Ramond charged black 3-brane
classical solution whose metric assumes the form\cite{Horowitz}:
\bea
\label{metric}
   ds^2 = &
H^{-1/2}(r)
    \left [ - f(r) (dx^0)^2 + 
(d x^i)^2 \right] \nonumber \\ +
 & H^{1/2}(r)
    \left [f^{-1} (r) dr^2 + r^2 d\Omega_{5}^2 \right ] \ ,
\eea
where $i=1,2,3$ and
$$   H(r)  = 1 + {L^4 \over r^4} \ , \qquad \  \  
f(r) = 1- {r_0^4\over
r^4}
\ .
$$
Here $d\Omega_5^2$ is the metric of a unit $5$ dimensional sphere,
${\bf S}^5$. 

In general, a $d$-dimensional sphere of radius $L$ 
may be defined by a constraint
\be
\sum_{i=1}^{d+1} (X^i)^2 = L^2
\ee
on $d+1$ real coordinates $X^i$. It is a positively curved 
maximally symmetric space with symmetry group $SO(d+1)$.
Similarly,
the $d$-dimensional Anti-de Sitter space, $AdS_d$,
is defined by a constraint
\be \label{embed}
(X^0)^2 + (X^d)^2 - \sum_{i=1}^{d-1} (X^i)^2 = L^2\ ,
\ee
where $L$ is its curvature radius.
$AdS_d$ is a negatively curved maximally symmetric space with
symmetry group $SO(2, d-2)$.
There exists a subspace of $AdS_d$ called the Poincar\' e wedge,
with the metric
\be \label{Poin}
ds^2 = {L^2 \over z^2} \left(dz^2 -(dx^0)^2 + \sum_{i=1}^{d-2}
(dx^i)^2\right)\ ,
\ee
where $z\in [0,\infty)$. 
In these coordinates the boundary of $AdS_d$ is at $z=0$.

The event horizon of the black 3-brane metric (\ref{metric})
is located at $r=r_0$.
In the extremal limit $r_0 \rightarrow 0$
the 3-brane metric becomes
\bea
\label{geom}
ds^2 = & \left (1+{L^4\over r^4}\right )^{-1/2}
\left (- (dx^0)^2 +(dx^i)^2 \right ) \nonumber \\
& + \left (1+{L^4\over r^4}\right )^{1/2}
\left ( dr^2 + r^2 d\Omega_5^2 \right )\ 
\ .
\eea
Just like the stack of parallel, ground state D3-branes, 
the extremal solution
preserves 16 of the 32 supersymmetries present 
in the type IIB theory. Introducing $z={L^2\over r}$,
one notes that the limiting form of (\ref{geom})
as $r\rightarrow 0$ 
factorizes into the direct product of
two smooth spaces, the Poincar\' e wedge (\ref{Poin})
of $AdS_5$, 
and ${\bf S}^5$,
with equal radii of curvature $L$. 
The 3-brane geometry may be thus
viewed as a semi-infinite throat of radius $L$ which for
$r \gg L$ opens up into flat $9+1$ dimensional space.
Thus, for $L$ much larger than the string length scale,
$\sqrt {\alpha'}$, 
the entire 3-brane geometry has small curvatures
everywhere and is appropriately described by the supergravity
approximation to type IIB string theory.

The relation between $L$
and $\sqrt{\alpha'}$ 
may be found by equating the gravitational
tension of the
extremal 3-brane classical solution to $N$ times the tension
of a single D3-brane, and one finds
\be\label{throatrel}
L^4 = g_{\rm YM}^2 N \alpha'^2\ .
\ee 
Thus, the size of the throat in string units is $\lambda^{1/4}$.
This remarkable emergence of the `t Hooft coupling
from gravitational considerations is at the heart of the success of
the AdS/CFT correspondence. Moreover,
the requirement $L\gg \sqrt{\alpha'}$ translates into
$\lambda \gg 1$: the gravitational approach is valid when
the `t Hooft coupling is very strong and the 
perturbative field
theoretic methods are not applicable.

\section{The AdS/CFT Correspondence}

Consideration of low-energy processes in the 
3-brane background\cite{absorption}
indicates that, in the low-energy limit, the 
$AdS_5 \times {\bf S}^5$ throat 
region ($r \ll L$) decouples from the asymptotically flat
large $r$ region. Similarly, the ${\cal N}=4$ supersymmetric 
$SU(N)$ gauge theory on the stack of $N$ D3-branes decouples 
in the low-energy limit from the bulk closed string theory.
Such considerations 
prompted Maldacena\cite{jthroat} to conjecture that type IIB string 
theory on $AdS_5 \times {\bf S}^5$, of radius $L$ given in
(\ref{throatrel}), is dual to
the ${\cal N}=4$ SYM theory. The number of colors in the gauge
theory, $N$, is dual to the number of flux units of the
5-form Ramond-Ramond field strength.

It was further conjectured in \cite{US,EW} that there
exists a one-to-one map between gauge invariant operators in the CFT and
fields (or extended objects) in AdS$_5$. The dimension $\Delta$ of an
operator is determined
by the mass of the dual field in AdS$_5$. For example, for scalar operators
one finds that $\Delta (\Delta-4)= m^2 L^2$. 
Precise methods
for calculating correlation functions
of various operators in a CFT using its dual formulation 
were also formulated\cite{US,EW}. 
They involve calculating the string theory path
integral as a function of the boundary conditions in AdS$_5$, which
are imposed near $z=0$.

If the number of colors $N$ is sent to infinity
while $g_{\rm YM}^2 N$ is held fixed and large, then there are small
string scale corrections to the supergravity limit\cite{jthroat,US,EW}
which proceed in powers of
${\alpha'\over L^2} = \left ( g_{\rm YM}^2 N \right)^{-1/2}
.
$
If we wish to study finite $N$, then there are also string loop
corrections in powers of
$ {\kappa^2\over L^8} \sim N^{-2}
.
$
As expected, taking $N$ to infinity enables us to take
the classical limit of the string theory on $AdS_5\times {\bf S}^5$.

Immediate support for the AdS/CFT correspondence comes from
symmetry considerations\cite{jthroat}. The isometry group of
$AdS_5$ is $SO(2,4)$, and this is also the conformal group in
$3+1$ dimensions. In addition we have the isometries of ${\bf S}^5$ which
form $SU(4)\sim SO(6)$. This group is identical to the R-symmetry of
the ${\cal N}=4$ SYM theory. After including the fermionic generators
required by supersymmetry, the full isometry supergroup of the
$AdS_5\times {\bf S}^5$ background is $SU(2,2|4)$, which is identical to
the ${\cal N}=4$ superconformal symmetry.

To formulate an AdS/CFT duality with a reduced amount of supersymmetry, 
we may place the stack of D3-branes at the tip of
a 6-dimensional Ricci flat cone,\cite{ks,Kehag,KW} 
whose base is a 5-dimensional compact Einstein space $Y_5$.
The metric of such a cone is $dr^2 + r^2 ds_Y^2$; therefore,
the 10-d metric produced by the D3-branes is obtained from (\ref{geom})
by replacing $d\Omega_5^2$, the metric on ${\bf S}^5$,
by $ds_Y^2$, the metric on $Y_5$. In the $r\to 0$ limit we then find
the space $AdS_5\times Y_5$ as the candidate dual of the CFT
on the D3-branes placed at the tip of the cone. 
The isometry group of $Y_5$ is smaller than $SO(6)$,
but $AdS_5$ is the ``universal'' factor present in the dual
description of any large $N$ CFT, making the $SO(2,4)$ conformal
symmetry geometric.

The fact that after the compactification on $Y_5$ 
the string theory is 5-dimensional supports earlier ideas on the necessity
of the 5-th dimension to describe 4-d gauge theories\cite{Polyakov}.
The $z$-direction is dual to the energy scale of the gauge theory: 
small $z$ corresponds to the UV domain of the gauge theory, 
while large $z$ to the IR. 

In the AdS/CFT duality, type IIB strings
are dual to the chromo-electric flux lines in the gauge theory, 
providing a string theoretic set-up for calculating 
the quark anti-quark potential\cite{Malda}.
The quark and anti-quark are placed near the boundary of Anti-de Sitter space 
($z=0$), and the fundamental string connecting them is required
to obey the equations of motion following from the Nambu action.
The string bends into the interior ($z>0$), and the maximum value
of the $z$-coordinate increases with the separation $r$ between quarks.
An explicit calculation of the string action gives an 
attractive $q\bar q$ potential\cite{Malda}: 
\be
V(r) = -\frac{4\pi^2 \sqrt{\lambda }}{\Gamma \left(\frac{1}{4} \right)^4 r} \ .
\ee
Its Coulombic $1/r$ dependence is required by the conformal invariance of 
the theory. Historically, a dual string description was hoped for mainly in
the cases of confining gauge theories, where long confining
flux tubes have string-like properties. In a pleasant surprise, we have seen that
a string description can be applicable to non-confining theories too,
due to the presence of extra dimensions in the string theory.

\subsection{
Spinning Strings vs. Long Operators}

A few years ago it was noted that the
AdS/CFT duality becomes particularly powerful when
applied to operators with large quantum numbers. One class of
such single-trace ``long operators'' are the BMN operators\cite{Berenstein}
that carry a large R-charge in the SYM theory and contain
a finite number of impurity insertions. The R-charge is dual
to a string angular momentum on the compact space $Y_5$.
So, in the BMN limit the relevant states are short closed strings
with a large angular momentum, and a small amount of vibrational excitation.
Furthermore, by increasing the number of impurities the string can be
turned into a large semi-classical object moving in $AdS_5\times Y_5$.
Comparing such objects with their dual long operators has become a very
fruitful area of research\cite{Tseytlinrev}. Work in this direction
has also produced a great deal of evidence that the ${\cal N}=4$
SYM theory is exactly integrable (see \cite{Beisert,Belitsky} for recent reviews).

A familiar example of a 
gauge theory operator with a large quantum number is a
twist-2 operator  carrying a large spin $J$,              
$  {\rm Tr}\ F_{+ \mu} D_+^{J-2} F_+^{\ \ \mu} $. In QCD, such operators
play an important role in studies of deep inelastic scattering\cite{Gross}.
In the ${\cal N}=4$ SYM theory, the dual of such a high-spin operator
is a folded string spinning around the center of $AdS_5$.\cite{Gubser:2002tv}
In general, for a high spin, 
the anomalous dimension of such an operator is \cite{Korchemsky}
\be
\Delta- (J+2)\rightarrow f(\lambda)\ln J
\ .\ee
Calculating the energy of the spinning folded string, we
find that the AdS/CFT prediction is\cite{Gubser:2002tv}
\be f(\lambda)\rightarrow {\sqrt{\lambda}\over \pi}
\ ,\ee
in the limit of large `t Hooft coupling. 
For small $\lambda$, perturbative calculations in the
large ${\cal N}=4$ SYM theory up to
3-loop order give\cite{Kotikov}
\be
f(\lambda)= {1\over 2\pi^2}
\left (\lambda- {\lambda^2\over 48} + {11 \lambda^3\over 11520}
+ O(\lambda^4)\right )
\ee
An approximate extrapolation formula, suggested
in \cite{Kotikov} works with about 10\% accuracy: 
\bea
\tilde f(\lambda)& ={12\over \pi^2} \left (-1+ \sqrt{1+\lambda/12}
\right )\nonumber \\& =
{1\over 2\pi^2}
\left (\lambda- {\lambda^2\over 48} + {\lambda^3\over 1152}
+ O(\lambda^4)\right )
\eea
Note that $\tilde f$ has a branch cut running from $-\infty$
to $-12$. Thus, the series has a finite radius of convergence,
in accord with general arguments about planar gauge theory given by
`t Hooft. The fact that the branch point is at a negative $\lambda$
suggests that in the ${\cal N}=4$ SYM theory
the perturbative series is alternating, and that there is no problem
in extrapolating from small to large $\lambda$ along the positive
real axis. It is, of course, highly desirable to find an exact formula for
$f(\lambda)$. Recent work\cite{Bern}
raises hopes that a solution of this problem
is within reach.

\section{Thermal Gauge Theory from Near-extremal D3-branes}

\subsection{Entropy}

An important black hole observable is the Bekenstein-Hawking (BH)
entropy, which is proportional to the area of the event horizon,
$S_{BH}= A_h/(4 G)$.
For the $3$-brane solution (\ref{metric}), 
the horizon is located at
$r=r_0$.
For $r_0>0$ the $3$-brane carries some excess
energy $E$ above its extremal value, and the
BH entropy is also non-vanishing. 
The Hawking temperature is then defined by
$ T^{-1} = \partial S_{BH}/\partial E$.

Setting $r_0\ll L$ in (\ref{geom}), we obtain
a near-extremal 3-brane geometry, whose 
Hawking temperature is found to be $T= r_0/(\pi L^2)$.
The small $r$ limit of this geometry is ${\bf S}^5$ times
a certain black hole in $AdS_5$.
The 8-dimensional ``area'' of the event horizon is
$ A_h = \pi^6 L^8 T^3 V_3 $,
where $V_3$ is the spatial volume of the D3-brane
(i.e. the volume of the $x^1, x^2, x^3$ coordinates).
Therefore, the BH entropy is\cite{gkp}
\begin{equation}
\label{bhe}
S_{BH}= {\pi^2\over 2} N^2 V_3 T^3
\ .
\end{equation}
This gravitational entropy of a near-extremal 3-brane
of Hawking temperature $T$ is to be identified
with the entropy
of ${\mathcal N}=4$ supersymmetric $U(N)$ gauge theory
(which lives on $N$ coincident D3-branes) heated up to
the same temperature.

The entropy of a free $U(N)$ ${\cal N}=4$
supermultiplet, which consists of the gauge field, $6 N^2$ massless
scalars and $4 N^2$ Weyl fermions, can be calculated using
the standard statistical mechanics of a massless gas (the black body 
problem), and the answer is
\be \label{ffc}
S_0= {2 \pi^2\over 3} N^2 V_3 T^3
\ .
\ee
It is remarkable that the 3-brane geometry
captures the $T^3$ scaling characteristic of a conformal
field theory (in a CFT this scaling is guaranteed by the extensivity of
the entropy and the absence of dimensionful parameters).
Also, the $N^2$ scaling indicates the presence of $O(N^2)$
unconfined degrees of freedom, which is exactly what we expect in
the ${\cal N}=4$ supersymmetric $U(N)$ gauge theory.
But what is the explanation of the relative factor of $3/4$ between $S_{BH}$ and
$S_0$?
In fact, this factor
is not a contradiction but rather a {\it prediction} about the strongly
coupled ${\cal N}=4$ SYM theory at finite temperature. 
As we argued above, the supergravity calculation of the
BH entropy, (\ref{bhe}),
is relevant to the $\lambda\rightarrow\infty$ limit of the 
${\cal N}=4$ $SU(N)$ gauge theory,
while the free field calculation, (\ref{ffc}), 
applies to the $\lambda\rightarrow
0$ limit. Thus, the relative factor of $3/4$ is not a discrepancy:
it relates two different limits of the theory. 
Indeed, on general field theoretic
grounds, in the `t Hooft large $N$
limit the entropy is given by\cite{GKT}
\be
S= {2 \pi^2\over 3} N^2 f(\lambda) V_3 T^3
\ .\ee
The function $f$ is certainly not constant: Feynman graph calculations
valid for small $\lambda=g_{\rm YM}^2 N$ give\cite{Foto}
\be \label{weak} 
f(\lambda) = 1 - {3\over 2\pi^2} \lambda 
+{3+\sqrt 2\over \pi^3} \lambda^{3/2} + \ldots
\ee
The BH entropy in supergravity, (\ref{bhe}),
is translated into the prediction that 
\be
\lim_{\lambda\rightarrow \infty}
f(\lambda ) = {3\over 4}\ .
\ee 
A string theoretic calculation of the leading correction
at large $\lambda$ gives\cite{GKT}
\be \label{strong} 
f(\lambda) = {3\over 4}  +  {45\over 32}\zeta(3)
\lambda^{-3/2} + \ldots
\ee
These results are consistent with a monotonic function
$f(\lambda)$ which decreases from 1 to 3/4 as $\lambda$ is increased
from 0 to $\infty$. The 1/4 deficit compared to the free field value
is a strong coupling effect predicted by the AdS/CFT correspondence.

It is interesting that similar deficits have been observed in 
lattice simulations of deconfined non-supersymmetric 
gauge theories\cite{Karsch,Gavai,Bringoltz}.
The ratio of entropy to its free field value,
calculated as a function of the temperature, is found to
level off at values around
$0.8$ for $T$ beyond 3 times the deconfinement temperature $T_c$.
This is often interpreted as the effect of a sizable coupling.
Indeed, for $T= 3 T_c$, the lattice estimates indicate
that $g_{\rm YM}^2 N \approx 7$.\cite{Gavai}
This challenges
an old prejudice that the QGP is inherently very weakly coupled.
We now turn to
calculations of the shear viscosity where strong coupling effects
are even more pronounced. 

\subsection{Shear Viscosity}

The shear viscosity $\eta$ may be read off from the form of
the stress-energy tensor 
in the local rest frame of the fluid where $T_{0i}=0$:
\be
T_{ij} = p \delta_{ij} -
\eta (\partial_i u_j + \partial_j u_i -{2\over 3} \delta_{ij}
\partial_k u_k )
\ ,
\ee
where $u_i$ is the 3-velocity field. 
The viscosity can be also determined\cite{Policastro}
through the Kubo formula
\be
\eta = \lim_{\omega\to 0} {1\over 2\omega}
\int dt d^3 x e^{i\omega t} \langle [T_{xy}(t,\vec x), T_{xy}(0,0)] \rangle
\ee
For the ${\cal N}=4$ supersymmetric YM theory this 
2-point function may be computed from absorption of a low-energy
graviton $h_{xy}$
by the 3-brane metric\cite{absorption}.
Using this method, it was found\cite{Policastro}
that at very strong coupling 
\be
\eta= {\pi\over 8} N^2 T^3
\ ,
\ee
which implies 
\be \label{bound}
{\eta\over s} = {\hbar \over 4\pi}
\ee
after 
$\hbar$ is restored in the calculation (here 
$s = S/V_3$ is the entropy density).
It has been proposed\cite{Kovtun}
that this value is the universal lower bound on
$\eta/s$.
Indeed, at weak coupling $\eta/s$ is very
large, $\sim {1\over \lambda^2 \ln (1/\lambda)}$,
and there is evidence that it 
decreases monotonically as the coupling is increased\cite{Buchel}.

The appearance of $\hbar$ in (\ref{bound})
is reasonable on general physical grounds\cite{Kovtun}. 
The shear viscosity $\eta$ is of order the 
energy density times quasi-particle mean free time $\tau$. 
So, $\eta/s $ is of order
of the energy of a quasi-particle times its mean free time, 
which is bounded from below by the uncertainty principle
to be some constant times $\hbar$.
The AdS/CFT correspondence fixes this constant to be $1/(4\pi)$,
which is not far from some earlier estimates\cite{Hirano}.

For known fluids (e.g. helium, nitrogen, water) $\eta/s$
 is considerably higher than the proposed lower bound\cite{Kovtun}.
On the other hand,
the Quark-Gluon Plasma produced at RHIC is believed  
to have a very low $\eta/s$, within a factor of 2 of 
the bound (\ref{bound})\cite{Teaney,Hirano}. 
This suggests that it is rather strongly
coupled.
Recently a new term, sQGP, which stands for 
``strongly coupled Quark-Gluon Plasma,''
has been coined to describe the deconfined state observed at 
RHIC\cite{Gyulassy,Shuryak} (a somewhat different term,
``Non-perturbative Quark-Gluon Plasma,'' was proposed in
\cite{Pisarsky}).
As we have reviewed, the AdS/CFT correspondence is a 
theoretical laboratory which allows
one to study analytically an extreme example
of such a new state of matter: the thermal
${\cal N}=4$ SYM theory at very strong `t Hooft coupling. 

In a CFT, the pressure is related to the energy density by $p=3e$. Hence, 
the speed of sound satisfies
$ c_s^2 = dp/de = {1\over 3}\ .
$
Recent lattice QCD calculations indicate that, while $c_s^2$ is much lower for 
temperatures slightly
above $T_c$, it gets close to $1/3$ for $T\geq 2 T_c$.\cite{Gavai}
Thus, for some range of temperatures starting around $2T_c$,
QCD may perhaps be treated as an approximately conformal, yet non-perturbative,
gauge theory.
This suggests that AdS/CFT methods could indeed be useful in studying 
the physics of sQGP,
and certainly gives strong motivation for continued experimental and
lattice research.

Lattice calculations indicate that the deconfinement temperature
$T_c$ is around $175\ MeV$, and  
the energy density is $\approx 0.7\ GeV/fm^3$,
around 6 times the nuclear energy density.
RHIC has reached energy densities around $14\ GeV/fm^3$, 
corresponding to $T \approx 2 T_c$.
Furthermore, in a few years, heavy ion collisions at the
LHC are expected to reach temperatures up to $5 T_c$.  
Thus, RHIC and LHC should provide a great deal of useful information
about the conjectured 
quasi-conformal temperature range of QCD.


\section{String Duals of Confining Theories}

It is possible to generalize the AdS/CFT correspondence in 
such a way that the quark anti-quark potential is linear at large distance.
In an effective 5-dimensional approach\cite{Polyakovnew}
the necessary metric is
\be \label{conffive}
ds^2 ={dz^2\over z^2} + a^2(z) \big (-(dx^0)^2 + (dx^i)^2 \big )
\ee
and the space must end at a maximum value of 
$z$ where the ``warp factor'' $a^2(z_{\rm max})$
is finite.\footnote{
A simple model of confinement\cite{PS} is obtained for $a(z) = 1/z$ in 
(\ref{conffive}), i.e. the metric is a slice of $AdS_5$ cut off at
$z_{\rm max}$. Hadron spectra in models of this type were studied in
\cite{Braga,Erlich,deTera}.}
Placing widely separated
probe quark and anti-quark near $z=0$,
we find that the string connecting them bends toward larger $z$ until
it stabilizes at $z_{\rm max}$ where its tension is minimized at the value
${a^2(z_{\rm max})\over 2\pi\alpha'}$.
Thus, the confining flux tube is described by a fundamental string
placed at $z= z_{\rm max}$ parallel to one of the $x^i$-directions.
This establishes a duality between ``emergent'' chromo-electric flux tubes
and fundamental strings in certain curved string theory backgrounds.

Several 10-dimensional supergravity
backgrounds dual to confining gauge theories are now known, but
they are somewhat more complicated than (\ref{conffive}) in that
the compact directions are ``mixed'' with the 5-d $(x^\mu,z)$ space.
Witten\cite{Wittenconf} constructed a background in the universality class of 
non-supersymmetric pure glue gauge theory. 
While in this background there is no asymptotic freedom in the UV, 
hence no dimensional transmutation, 
the background has served as a simple model of confinement
where many infrared observables have been calculated
using the classical supergravity.
For example,
the lightest glueball masses have been 
found from normalizable fluctuations around the 
supergravity solution\cite{Csaki}. 
Their spectrum is discrete, and resembles  
qualitatively the results of lattice simulations in the pure glue theory.

Introduction of a minimal (${\cal N}=1$) supersymmetry 
facilitates construction of gauge/string dualities. As discussed earlier,
a useful method is to place a stack of D-branes at the tip 
of a six-dimensional
cone, whose base is $Y_5$. For $N$ D3-branes, one finds the
background $AdS_5\times Y_5$ dual to a superconformal gauge theory.
Furthermore,
there exists an interesting way of breaking the conformal invariance
for spaces $Y_5$ whose topology includes an ${\bf S}^2$ factor.
At the tip of the cone over
$Y$ one may add $M$ wrapped D5-branes to the $N$ D3-branes.
The gauge theory on such a combined stack is no longer conformal;
it exhibits a novel pattern of quasi-periodic renormalization group flow,
called a duality cascade\cite{KT,KS} (for reviews, see
\cite{Herzog,Strassler}).

To date, the most extensive study of a theory of this type has been carried out
for a simple 6-d cone called the
conifold, where one finds a ${\cal N}=1$ supersymmetric
$SU(N) \times SU(N+M)$ theory coupled to chiral superfields  
$A_1, A_2$ 
in the $ ({\bf N},\overline{\bf{N+M}})$ representation, and
$B_1, B_2$ 
in the $ (\overline{{\bf N}},\bf{N+M})$ representation.
In type IIB string theory, D5-branes source the 7-form field
strength from the Ramond-Ramond sector, which is Hodge dual to
the 3-form field strength. Therefore, 
the $M$ wrapped D5-branes
create $M$ flux units of this field strength
 through the 3-cycle in the conifold;
this number 
is dual to the difference between the numbers of colors 
in the two gauge groups. 
An exact
non-singular supergravity solution dual to this gauge theory,
incorporating the 3-form and the
5-form R-R field strengths, and their back-reaction on the geometry,  
has been found\cite{KS}.
This back-reaction creates a 
``geometric transition'' to the deformed conifold
\be \label{defconi}
\sum_{a=1}^4 z_a^2 = \epsilon^2
\ ,
\ee
and introduces a ``warp factor'' so that the full 10-d geometry has
the form
\be \label{specans}
ds^2 =   h^{-1/2}(\tau) \left (- (dx^0)^2 +  (d x^i)^2 \right ) 
 +  h^{1/2}(\tau) d\tilde s_6^2 \ ,
\ee
where $d\tilde s_6^2$ is the Calabi-Yau metric of the deformed conifold,
which is known explicitly.

The field theoretic interpretation of this solution is unconventional.
After a finite amount of RG flow, the
$SU(N+M)$ group undergoes a Seiberg
duality transformation\cite{Seiberg}. After this transformation,
and an interchange of the two gauge groups,
the new gauge theory is $SU(\tilde N )\times SU(\tilde N+ M)$
with the same matter and superpotential, and with $\tilde N=N-M$.
The self-similar structure of the gauge theory under the Seiberg
duality is the crucial fact that allows this pattern to repeat many times.
If $N= (k+1) M$, where $k$ is an integer, then the 
duality cascade stops after $k$
steps, and we find a $SU(M)\times SU(2M)$ gauge theory. This IR
gauge theory exhibits a multitude of interesting effects visible
in the dual supergravity background. One of them is 
confinement,
which follows from the fact that the warp factor $h$ is finite and
non-vanishing at the smallest radial coordinate,
$\tau=0$, which roughly corresponds to $z=z_{\rm max}$ in an
effective 5-d approach (\ref{conffive}).
This implies that the quark anti-quark potential grows linearly
at large distances.
Other notable IR effects are 
chiral symmetry breaking, and the Goldstone mechanism\cite{GHK}.
Particularly interesting
is the appearance of an entire ``baryonic branch'' of the moduli space
in the gauge theory, whose existence has been recently demonstrated
also in the dual supergravity language\cite{Butti}.

Besides providing various new insights into the IR physics of confining
gauge theories,
the availability of their string duals enables one to study 
Deep-Inelastic and hadron-hadron scattering in this new language\cite{PS}.

\section{Summary}

Throughout its history, string theory has been intertwined with 
the theory of strong interactions. The AdS/CFT 
correspondence\cite{jthroat,US,EW} succeeded
in making precise connections between conformal 4-dimensional
gauge theories and
superstring theories in 10 dimensions. This duality leads to a multitude
of dynamical predictions about strongly coupled gauge theories.
When extended to theories at finite temperature, it serves
as a theoretical laboratory for studying a novel state of matter:
a gluonic plasma at very strong coupling. This appears to have
surprising connections to the new state of matter, sQGP, observed at 
RHIC\cite{Adcox}.

Extensions of the AdS/CFT correspondence to confining gauge theories
provide new geometrical viewpoints on such important phenomena as
chiral symmetry breaking and dimensional transmutation. They allow for studying
meson and glueball spectra, and high-energy scattering, in model
gauge theories.

This recent progress offers new tantalizing hopes that an analytic
approximation to QCD will be achieved along this route, at least
for a large number of colors. But there is much work that remains to
be done if this hope is to become reality: understanding the string
duals of weakly coupled gauge theories remains an important open problem.

\section*{Acknowledgments}
I am grateful to the organizers of the 2005
Lepton-Photon Symposium in Uppsala for giving me an opportunity to present
this talk in a pleasant and stimulating environment.
I also thank Chris Herzog for his very useful input.
This research is
supported in part by the National Science Foundation Grant No.~PHY-0243680.
Any opinions, findings, and conclusions or recommendations expressed in
this material are those of the authors and do not necessarily reflect
the views of the National Science Foundation.

\end{document}